%% file: main.tex
\documentclass[%
 reprint,
 amsmath,amssymb,
 aps,
prl 
]{revtex4-2}

\usepackage{aas_macros}
\usepackage{graphicx}%
\graphicspath{{./}{figures/}}
\usepackage{dcolumn}%
\usepackage{bm}%
\usepackage[usenames,dvipsnames]{xcolor}
\usepackage{hyperref}%
\hypersetup{
	colorlinks=true,
	citecolor=NavyBlue,
	linkcolor=NavyBlue,
	urlcolor=NavyBlue,
}

\usepackage{epsfig}
\usepackage{url}
\usepackage[normalem]{ulem}
\usepackage{latexsym}
\usepackage{epsfig}
\usepackage{amsmath}
\usepackage{amssymb}
\usepackage{wasysym}
\usepackage{graphicx}
\usepackage{verbatim}
\usepackage{enumerate,mdwlist}%
\usepackage[titletoc]{appendix}
\usepackage{amsfonts}
\usepackage{pgfplots}
\usepackage[export]{adjustbox}
\usepackage{bbold}

\input{new_commands}

\newcommand{\lettersection}[1]{\textbf{\emph{#1}}.---}
\newcommand{\this}{Letter~}

\usepackage[acronym]{glossaries}
\newacronym{GW}{GW}{gravitational wave}
\newacronym{EM}{EM}{electromagnetic} %
\newacronym{GL}{GL}{gravitational lensing}
\newacronym{GO}{GO}{geometric optics} %
\newacronym{WO}{WO}{wave optics}
\newacronym{CBC}{CBC}{compact binary coalescence}
\newacronym{BBH}{BBH}{binary black hole}
\newacronym{BNS}{BNS}{binary neutron star}
\newacronym{NSBH}{NSBH}{neutron-star black-hole binary}
\newacronym{SIE}{SIE}{singular isothermal ellipsoid}
\newacronym{SNR}{SNR}{signal-to-noise ratio}
\newacronym{PE}{PE}{parameter estimation}

\makeatletter
\newcommand*{\glsplainhyperlink}[2]{%
  \colorlet{currenttext}{.}%
  \colorlet{currentlink}{\@linkcolor}%
  \hypersetup{linkcolor=currenttext}%
  \hyperlink{#1}{#2}%
  \hypersetup{linkcolor=currentlink}%
}
\let\@glslink\glsplainhyperlink
\makeatother

\begin{document}

\preprint{APS/123-QED}

\title{Observational Signatures of Highly Magnified Gravitational Waves from Compact Binary Coalescence}

\author{Rico K.~L.~Lo}
\email{kalok.lo@nbi.ku.dk}
\author{Luka Vujeva}
\author{Jose Mar\'ia Ezquiaga}
\author{Juno C.~L.~Chan}
\affiliation{Center of Gravity, Niels Bohr Institute, Blegdamsvej 17, 2100 Copenhagen, Denmark}

\date{\today}%

\begin{abstract}
Gravitational lensing has empowered telescopes to discover astronomical objects that are otherwise out of reach without being highly magnified by foreground structures.
While we expect \glspl{GW} from compact binary coalescences to also experience lensing, the phenomenology of highly magnified \glspl{GW} has not been fully exploited. In this Letter, we fill this gap and explore the observational signatures of these highly magnified \glspl{GW}.
We find that these signatures are robust against modeling details and can be used as smoking-gun evidence to confirm the detection of lensing of \glspl{GW} without any electromagnetic observation.
Additionally, diffraction becomes important in some cases, which limits the maximum possible magnification and gives waveform signatures of lensing that can only be observed by \gls{GW} detectors.
Even with current-generation observatories, we are already sensitive to these {rare,} highly magnified \glspl{GW} and {could} use them to probe the high-redshift Universe beyond the usual horizon.
\end{abstract}

\maketitle

\glsresetall
\lettersection{Introduction}
Our knowledge of the Universe is limited by the sensitivity of our current observatories. 
However, the cosmos has an aid for us, as the largest structures in the Universe, such as galaxies or clusters of galaxies, act as huge magnifying glasses to uncover objects %
at much farther distances. %
Thanks to this \gls{GL} effect, we have been able to observe the farthest galaxies at redshifts of $z \approx 13$ \citep{2023ApJ...957L..34W} and even single stars at $z \approx 6$ \citep{Welch_2022}. %
In the latter case, the star's light was amplified by a magnification of %
more than $\mu \sim 10^4-10^5$ times. 
\gls{GL} therefore offers a unique opportunity to probe the unexplored distant Universe with sources that would otherwise be too faint to detect. 

In the quest of probing the high-redshift Universe, \glspl{GW} from \glspl{CBC} %
play a special role as these signals travel essentially unaltered from the source to our detectors, keeping a pristine record of their astrophysical origin, gravitational interactions and cosmological environment.
\gls{GL} may magnify not only static but also transient sources. This has already been observed for supernovae \citep{2024SSRv..220...13S}. %
Within the current observations from the LIGO \cite{LIGOScientific:2014pky}, Virgo \cite{VIRGO:2014yos} and KAGRA \cite{KAGRA:2020tym} detectors, there is no evidence of lensing yet \cite{Hannuksela:2019kle,LIGOScientific:2021izm,Janquart:2023mvf,LIGOScientific:2023bwz}. 
This is expected given the rarity of \glspl{CBC}, $\approx 100\,/\text{yr}/\text{Gpc}^3$, and lensing itself, with an optical depth $\tau(z \sim 1) \lesssim 10^{-3}$ \cite{Oguri:2019fix}. 
Nonetheless, for high magnifications, \gls{GW} detectors are not flux limited and can probe the population of \glspl{CBC} across all redshifts.

Observing a lensed \gls{GW} will open a new frontier in \gls{GL}. 
It will entail the first detection of \gls{GL} on a cosmic messenger other than \gls{EM} radiation, probing new aspects of gravity. 
It will also constrain the population of compact binaries at high redshift \cite{Xu:2021bfn, Lo:2021nae} and enable other science, such as measuring the cosmic expansion \cite{Oguri:2019fix} and mapping the dark matter substructures~\cite{Tambalo:2022wlm,Caliskan:2023zqm}. 
Moreover, given the large wavelengths of \glspl{GW}, it may probe \gls{GL} in a new regime, \gls{WO}, where the waves deflected by the lens interfere with each other.
In this \this we explore the observational prospects of highly magnified \glspl{GW} and demonstrate that there are ``smoking-gun'' signatures for their detection. 
{Other smoking-gun signatures for GW lensing include waveform distortions due to type-II phase shifts~\cite{Dai:2017huk,Ezquiaga:2020gdt, Wang:2021kzt} and diffraction by compact lenses~\cite{Nakamura:1999uwi}.}

\lettersection{Highly magnified gravitational waves}
When the wavelength of a \gls{GW} $\lambda_{\text{GW}}$ is much smaller than the characteristic size of the lens $R_{\text{lens}} \gg \lambda_{\text{GW}}$, \gls{GO} is sufficient for describing the \gls{GL} of \glspl{GW}. The phenomenology is almost identical to the \gls{GL} of light \footnote{Note, however, we will discuss an exception later in this \this where \acrlong{GO} breaks down and a full \acrlong{WO} treatment is needed.}.
In this regime, when the arrival of a \gls{GW} can be distinctively identified in a one-dimensional time series of strain $h$ (for example, a \gls{GW} signal from a \gls{CBC} in a ground-based \gls{GW} detector), lensed \glspl{GW} coming from the same source are registered as multiple \gls{GW} triggers arriving at different times with different apparent luminosity distances $d_{\text{L}}^{\text{app}} = d_{\text{L}}^{\text{true}}/\sqrt{\mu}$, which will be smaller than the true luminosity distance of the source $d_{\text{L}}^{\text{true}}$ if a \gls{GW} signal is magnified ($\mu > 1$) %
and vice versa if demagnified. 
Each trigger also carries a distinctive phase shift encoding the reflection parity. %

This is similar to the \gls{GL} of light where separate images of the same source (when the images are resolvable) appear on a two-dimensional image plane $\vec{x} = (x_1, x_2)$ in the sky. 
Therefore, in the literature, we often use the term ``lensed images'' interchangeably with ``lensed \gls{GW} signals.'' 
One important distinction, though, is that, with \gls{GW} detectors, we can observe the phase evolution of a signal. Thus, we can observe waveform modifications due to \gls{GL} and not only changes in intensities. %

The magnification $\mu$ and the time delay $\Delta t$ of a lensed \gls{GW} signal relative to the unlensed one depend on the alignment between the source, the lens, and the observer.
Suppose a source is located at  $\vec{y} = (y_1, y_2)$ on another two-dimensional plane called the source plane; 
the lens equation (schematically, $\vec{y} = \vec{x} - \vec{\alpha}$) maps the possibly multiple image locations $\vec{x}$ to the source location $\vec{y}$ and $\vec{\alpha}$ is the deflection angle from \gls{GL}.

For some special locations on the source plane called caustics, %
images with formally infinite magnification can be formed on the image plane, though in reality, they have finite magnifications instead. 
These ``infinitely bright'' images trace out the critical curves 
on the image plane, and their corresponding caustics trace out boundaries of regions having different image multiplicity on the source plane \cite{Blandford:1986zz, Schneider:1992bmb}.
At these points, the mapping between the source and the image plane is singular.
The morphology of caustics can be very diverse and reflects the specifics of a lens. However, those that are stable on two-dimensional planes, and thus relevant here, are either a fold or a cusp caustic \cite{WH55, Schneider:1992bmb}.
The most common and simplest of those is a fold caustic, which consists of a collection of isolated singularity points \cite{Berry_Upstill}.
A cusp caustic forms when two fold caustics meet \cite{Berry_Upstill}, and thus it is astrophysically rarer than a fold caustic in the context of \gls{GL}.

When a source is located close to a fold caustic and inside a region where multiple images are formed, two images of the source with high magnifications will be formed to the left and the right of the corresponding critical curve, respectively. Moreover, these two images will have similar magnifications (equivalently, a relative magnification $\murel$ that is close to unity) but opposite parity \setcounter{footnote}{10}\footnote{The phenomenology for a source near a cusp caustic is similar to that near a fold caustic. The images formed close the corresponding critical curve will also be highly magnified and have very short time delays.
However, for a cusp, there will be three images (instead of two) near the critical curve. Moreover, two out of the three image pairs will have a relative magnification $\murel = 1/2$ with opposite parity, and the remaining pair will have a unit relative magnification but with the same parity \cite{Schneider:1992bmb}.}.
If the source is time varying, then the time delay between those two images will be short; therefore, the two signals may overlap in the time domain. Note that these qualitative features are insensitive to how one models a lens and universal whenever a source is located near a fold caustic \footnote{If the lens model includes sub-structures such as dark matter sub-halos and stars, then the relative magnification might deviate slightly from unity, e.g. see Refs.~\cite{Oguri:2017ock, Williams:2023jiq}, and the time delay remains small.}.
If we parametrize the perpendicular distance from the source to the fold caustic as $\Delta \theta_{\text{S}} \ll 1$, then as the source approaches the caustic~\cite{Schneider:1992bmb}, 
\begin{equation}
\label{eq:magnification_scaling}
\mu_{\pm} \sim  1/\sqrt{\Delta \theta_{\text{S}}} \sim \Delta t^{-1/3}\,,
\end{equation}
where $\mu_{\pm}$ are the magnification of the image near the critical curve with a positive and a negative parity, respectively. The relative magnification $\murel \equiv \mu_{+}/\mu_{-}$ has a correction from unity that scales as $\murel - 1 \sim \sqrt{\Delta \theta_{\text{S}}} \sim \Delta t^{1/3}$. Therefore, as the time delay between the two images shortens, their relative magnification becomes closer to one.

\begin{figure*}[t!]
    \centering
    \includegraphics[width=0.9\textwidth]{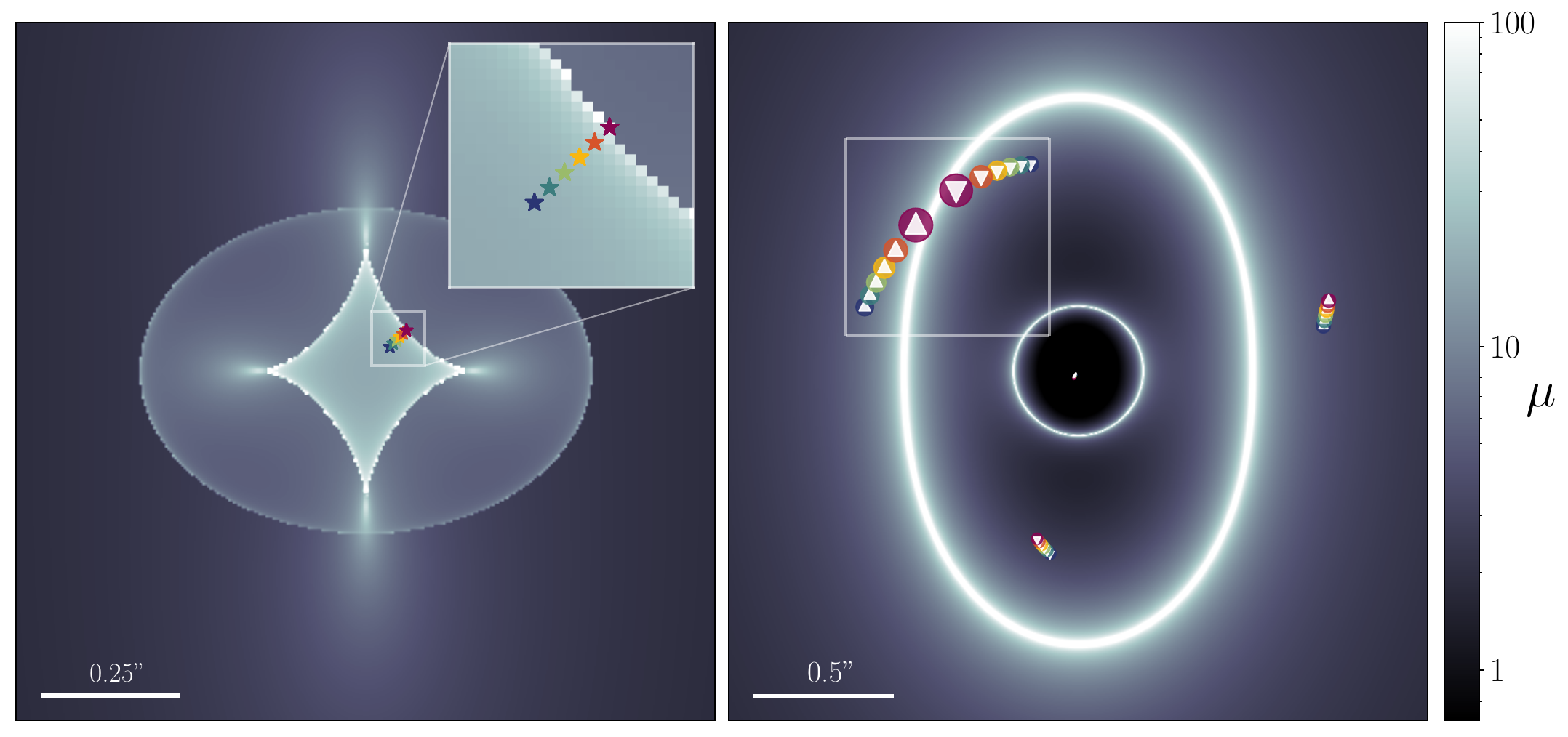}

    \caption{Point sources (represented by stars) placed 
    near the fold caustic in the source plane (left) of a galaxy lens with a cored singular isothermal ellipsoid density profile and their corresponding images (circles) with the same color
    in the image plane (right), shown over the magnification map $\mu$ of each plane, respectively.
    The size of the circles is determined by the magnification of each image. The triangle within each circle [upward-pointing triangles (downward-pointing triangles)] denotes the positive (negative) parity of the image. %
   Sources close to the fold caustic produce locally two images (white square frames) of opposite parity and similar high magnification on each side of the critical curve.
    }
    \label{fig:simulation_lens}
\end{figure*}

We illustrate this phenomenology
in Fig.~\ref{fig:simulation_lens}, where we construct a simple yet realistic galaxy lens system with a cored \acrlong{SIE} density profile \cite{kormann1994} of velocity dispersion $\sigma_v = 220$ km s$^{-1}$ with an ellipticity of $e=0.3$, a core of $r_{\text{c}}=0.1 ''$, and an external shear of $\gamma_{\text{ext}}=0.1$ at a redshift of $\zl = 0.5$ \footnote{Here we are using a spatially-flat $\Lambda$CDM model with cosmological parameters taken from Ref.~\cite{Planck:2018vyg}.}.
Point sources (at a redshift of $\zs=2$) are placed near the fold caustic on the internal side of the caustic. The image properties are calculated numerically using the lens modeling software \texttt{glafic} \cite{glafic2010, glafic2010software, glafic2021}. We only consider the two most magnified images relevant to this study (i.e., the images in the right panel inset in Fig.~\ref{fig:simulation_lens}). 
As expected, the relative magnification of the images $\murel$ rapidly approaches unity as the source position approaches the caustic, and the magnification $\mu$ of both images diverges.
The magnification of each image pair indeed follows the scaling as Eq.~\eqref{eq:magnification_scaling}.
In particular, the pair of images near the critical curve in Fig.~\ref{fig:simulation_lens} with the corresponding source located at $\Delta \theta_{\text{S}} \approx 3.6 \,\mu \text{arcseconds}$ away from the caustic would each have a magnification of $\mu \sim 10^3$, a relative magnification $|\murel - 1| \approx 6 \times 10^{-2}$, and a time delay $\Delta t \sim 10^{-2}$ s for the pair.

In this limit where a source is located near a fold caustic, the lensed gravitational waveform $\tilde{h}^{\text{L}}(f) = F(f) \tilde{h}^{\text{U}}(f)$ \cite{Nakamura:1999uwi} and the unlensed waveform $\tilde{h}^{\text{U}}(f)$ in the Fourier domain is related by the amplification factor $F(f)$ given by
\begin{equation}
\label{eq:F_two_images}
F(f { > 0}) = {\sqrt{\mu_1}e^{2\pi i f t_1}}\left[ 1 + \sqrt{\murel} \exp \left( 2\pi i f \Delta t - i\pi/2 \right)\right],
\end{equation}
where $\Delta t > 0$ is {effectively} the only free parameter here {since} $\murel \approx 1${ and} 
the magnification {$\mu_{1}$} and the time delay {$t_{1}$} of the first signal %
are absorbed by the apparent luminosity distance $d_{\text{L}}^{\text{app}}$ and the arrival time, respectively.
The lensed \gls{GW} signal that arrives slightly later must have an extra phase shift of $-\pi/2$ relative to the first signal \footnote{Equivalently, this means the first lensed signal out of the two highly magnified ones must be type I and the second one is type II.}.
These are very specific observational signatures for highly magnified \glspl{GW} that we can look for in \gls{GW} data. Note that, however, the high magnification does \emph{not} necessarily imply a loud \gls{GW} signal at a detector since the original unlensed signal could have been too quiet (e.g., too far away) for current-generation detectors to detect without the magnification from \gls{GL}. 

\lettersection{Overlapping GW signals from CBC} %
When the time delay $\Delta t$ between two \gls{GW} signals is less than their chirp time $\tau_{\text{chirp}}(f_{\text{low}}, f_{\text{high}}) \equiv \int_{f_{\text{low}}}^{f_{\text{high}}} dt/df \, df$, which is the time elapsed for the frequency of the GW $f$ to evolve from $f_{\text{low}}$ to $f_{\text{high}}$ (for a $30M_{\odot}-30M_{\odot}$ binary, $\tau_{\text{chirp}} \approx 1\,$s with $f_{\text{low}} = 20\,$Hz) \cite{Allen:2005fk}, the two signals overlap in time and interfere each other.
One might intuitively expect that matched-filter-based search pipelines that were designed to look for vanilla nonoverlapping \gls{GW} signals will fail here since the resultant linear superposition of the signals looks very distorted compared to the constituent waveforms.

However, this is actually not the case. In matched filtering, signal templates are slid across the noisy search data to compute the \gls{SNR} $\rho(t)$ as a function of the time lag $t$ between the template and the data \cite{Finn:1992wt, Allen:2005fk}.
The presence of a chirping signal (such as \gls{GW} signals from \glspl{CBC}) in the data produces a peak in the \gls{SNR} time series when an appropriate template is used.
Therefore, the two constituent waveforms will produce two peaks in the \gls{SNR} time series separated by $\Delta t$.

The ability of a \gls{GW} detector to resolve distinct peaks in a \gls{SNR} time series, meaning that one can differentiate the arrival of simply one signal or multiple signals separated by some (small) time delays, is analogous to the resolvability of a telescope to tell whether there is just a single object or multiple objects in an image. 
The time-delay resolution is thus determined by the width of those \gls{SNR} time series peaks, which is given by the full width at half maximum of the autocorrelation function of the template $\tau_{\text{autocorr}}$ (independent of the value of a \gls{SNR} peak).
For example, the waveform of a $30M_{\odot} - 30M_{\odot}$ \gls{BBH} merger \footnote{Note that these component masses (and all the masses mentioned in this \this hereafter except otherwise specified) are specified in the ``detector frame'', i.e., they are redshifted by a factor of $(1 + \zl)$ compared to their values in the ``source frame''.}, $\tau_{\text{autocorr}} \approx 3\times10^{-3}$ s (which scales 
{with the frequency at which the binary merges as $1/f_{\text{merger}}$ when it is within the detection band~\cite{Fairhurst:2010is}}%
). This means that the resolution is actually limited by the self-similarity of waveforms and not by the hardware limitation of a finite sampling rate $f_{\text{samp}}$ [for ground-based detectors, this gives a constraint of $1/(f_{\text{samp}}/2) \approx 3 \times 10^{-5}\,\text{s}\ll \tau_{\text{autocorr}}$].

\begin{figure}[t!]
    \centering
    \includegraphics[width=0.5\textwidth]{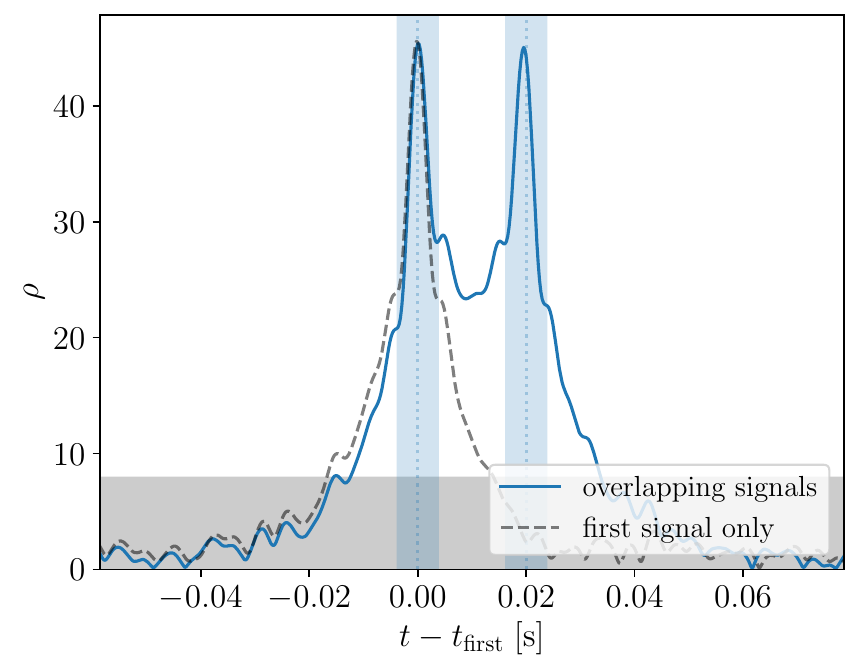}
    \caption{The \gls{SNR} $\rho$ computed using a template for a $30M_{\odot} - 30M_{\odot}$ \gls{BBH} merger on simulated data injected with two overlapping $30M_{\odot} - 30M_{\odot}$ \gls{BBH} signals separated by $\Delta t = 0.02$ s with a unit relative magnification and opposite parity (solid curve) and with only one \gls{BBH} signal (dashed curve).
    One can clearly identify two peaks above the fiducial detection threshold of $\rho \geq 8$ in the \gls{SNR} time series. The two peaks are $0.02$ s apart with roughly the same value of \gls{SNR} as expected.
    The full width at half maximum of the template's autocorrelation function $\tau_{\text{autocorr}}$ is indicated by the width of the vertical bands around the two peaks, which match the width of the \gls{SNR} peaks.
  	}
    \label{fig:snr_timeseries}
\end{figure}

Our discussion above applies naturally when the time delay from \gls{GL} near a fold caustic is smaller than the chirp time of the lensed \gls{GW} signal \footnote{The discussion is actually also applicable to a cusp caustic since the time delay is also small. However, note that the image properties, especially the relative magnification and parity, from a source near a cusp differ slightly from that near a fold caustic \cite{Note11}.}. This is illustrated in Fig.~\ref{fig:snr_timeseries}, where we show the \gls{SNR} time series computed using a template for a $30M_{\odot} - 30M_{\odot}$ \gls{BBH} merger on simulated strain data (with Gaussian noise at the Advanced LIGO design sensitivity \cite{design_sensitivity_curves_ligo}) injected with two overlapping $30M_{\odot} - 30M_{\odot}$ \gls{BBH} signals at an apparent luminosity distance $d_{\text{L}}^{\text{app}} = 490$ Mpc (akin to GW150914 \cite{LIGOScientific:2016aoc}), separated by a time delay $\Delta t = 0.02$ s with a relative magnification $\murel = 1$ \footnote{This is consistent with the scenario where the merger is at a redshift $\zl = 2$, and the two lensed \gls{GW} signals have a magnification of $\mu = 1000$ each, as in the setup of Fig.~\ref{fig:simulation_lens}.} as the solid curve and the same simulated data but with only one \gls{BBH} signal injected as the dashed curve, respectively.

As expected, there are two peaks in the \gls{SNR} time series with similar values ($\rho \approx 48$ \footnote{While it is certainly loud, this is not inconceivable nor an outlier, since this is essentially what GW150914 \cite{LIGOScientific:2016aoc} will look like in Advanced LIGO with design sensitivity in the future.}) separated exactly by the time delay $\Delta t$ even when the template consists of only one merger for the overlapping case.
Although the mismatch $\epsilon$ \cite{Owen:1995tm} between the superimposed waveform and the constituent waveform, $\epsilon \approx 30\%$ in this case, is not small, the value of the \gls{SNR} peaks is actually identical to the case where there is only one merger injected in the simulated data and not reduced to $\rho \sqrt{1-\epsilon} \approx 40$.
Therefore, one can get a rough estimate of $\mu_{\mathrm{rel}}$ and $\Delta t$ by simply reading off from the peaks of a \gls{SNR} time series for lensed \glspl{GW} near a critical curve.
Additionally, we can obtain the phase difference of the template at those peaks and explicitly check whether the recovered phase difference matches with the theoretical expectation~\footnote{If the template contains only the quadrupole gravitational radiation (which is usually the case \cite{Usman:2015kfa, Adams:2015ulm, Messick:2016aqy}), then the expected phase difference for this mode is $-\pi/4$ \cite{Dai:2017huk,Ezquiaga:2020gdt}.}.
From this, we see that a matched-filter-based \gls{CBC} search pipeline will be able to search for these overlapping \glspl{GW} without constructing a new bank of templates \footnote{We have explicitly verified that all these are still true even at a ``typical'' \gls{SNR} by replacing the injected \gls{GW} signals in the simulated data with much fainter ones, at an apparent luminosity distance $d_{\text{L}}^{\text{app}} = 2000$ Mpc instead of $490$ Mpc, such that the \gls{SNR} peak is about 12 instead.}.

Since there is a large mismatch when a nonlensed signal is matched with overlapping lensed signals, it implies that, if one were to carry out \gls{PE} on such an event, one might get a biased inference on the parameters.
We verify this by performing \glspl{PE} on the aforementioned simulated data injected with overlapping \gls{GW} signals using a waveform model that contains one and two mergers, respectively.
We find that most of the parameters are indeed biased when the overlapping is left unaccounted for and that we can correctly recover all of the waveform parameters, the time delay $\Delta t$ and the $-\pi/2$ phase shift from lensing after taking the overlapping into account \footnote{This is in line with the literature, e.g.~Ref.~\cite{Pizzati:2021apa}, that current \gls{GW} data analysis tools such as search pipelines and \gls{PE} codes are capable of handling overlapping signals, once a search pipeline correctly identifies the time delay between signals.}.

The fact that two signals are found by the same search template with roughly the same \gls{SNR}{, dephased by $\pi/2$, and} arriving only $\Delta t \ll 1$ s apart is smoking-gun evidence that the signals are lensed.
The chance coincidence of two nonlensed \glspl{GW} arriving $\Delta t \ll \tau_{\text{chirp}}$ apart %
is extremely small. Given the average number of \gls{GW} detections per unit time $R$ at a given sensitivity, we can compute a \emph{conservative upper bound} on the false-alarm probability $p_{\text{FA}}$ per event by assuming that the arrival of nonlensed \glspl{GW} is a Poisson process as
\begin{equation}
	p_{\text{FA}} < 1 - e^{-R\Delta t} \approx 5.79 \times 10^{-8} \left( \dfrac{R}{1/2 \,\text{day}^{-1}} \right) \left( \dfrac{\Delta t}{10^{-2}\,\text{s}} \right),
\end{equation}
when $R\Delta t \ll 1$. 
The false-alarm probability will be \emph{even smaller} after imposing the requirement on matching the same template with the same \gls{SNR} {and phase shifted by $\pi/2$ \footnote{{Suppose $\sim0.01\%$ of all events lensed by galaxies have $\mu>100$~\cite{Oguri:2018muv} and an optical depth of $\sim10^{-3}$, then the rate of observing a highly magnified event in a year would be $\sim10^{-5}(R/100\,\mathrm{yr}^{-1})$. This rough estimate does not take into account the selection bias towards high magnifications and the possibility that groups and clusters of galaxies may lead to more highly magnified events \cite{Vujeva:2025kko}. While $R$ can be $\sim 10^5\,\mathrm{yr}^{-1}$ for future-generation detectors, the number of signals having their time of arrival (not requiring matching the search template, etc, as mentioned in the main text) being within $\tau_{\mathrm{autocorr}} \approx 10^{-3}\,\mathrm{s}$ is only about unity \cite{Pizzati:2021apa}.}}}. 

\lettersection{Observing diffraction in gravitational-wave lensing} 
Within the \gls{GO} approximation, the magnification of an image diverges when it gets closer to a caustic as $\Delta \theta_{\text{S}} \to 0$ [cf.~Eq.~\eqref{eq:magnification_scaling}].
This mathematical divergence does not occur in nature because the finite size of the source or the finite wavelength will eventually regularize the magnification.
In the context of \glspl{GW} from \glspl{CBC},
its wave nature will limit the maximum possible magnification in \gls{GL}.

In the weak-field gravity limit, \gls{GL} amounts to solving the wave propagation over a slightly perturbed flat spacetime.
The wave equation can be solved using a Kirchoff diffraction integral that defines the amplification factor $F(f)$ in the Fourier domain \cite{Schneider:1992bmb, Nakamura:1999uwi}.
With the thin-lens approximation, the diffraction integral when the source is located at $\vec{y}$ on the source plane is given by $F(f) = \frac{w}{2\pi i}\int d^2x\, e^{iw \phi(\vec x,\vec y)}$,
where $\phi$ is the Fermat potential that is related to the time delay and that we introduce a dimensionless frequency $w \equiv (1+\zl)\left(D_{\text{L}}D_{\text{S}}/cD_{\text{LS}}\right)\theta_{\star}^2 2\pi f$ with $\theta_{\star}$ being the angular scale of the lens and $D_{\text{L, S, LS}}$ being the angular diameter distance from the observer to the lens, to the source and between the lens and the source, respectively.

The wave propagation around a fold caustic can be described by first expanding the Fermat potential up to cubic order \cite{Blandford:1986zz, Schneider:1992bmb, Berry_Upstill} as $\phi \approx \frac{\phi_{11}}{2}x_1^2-x_1y_1-x_2y_2+\frac{\phi_{222}}{6}x_2^3$,
where $\phi_{i \dots j} \equiv \partial_{x_j} \dots \partial_{x_i} \phi$ denotes the partial derivative of the Fermat potential $\phi$ with respect to $x_i \dots x_j$ ($i,j$ are indices taking on values $\left\{1,2\right\}$) evaluated at the caustic.
With this ansatz, we find an analytic solution for the diffraction integral:
\begin{multline}
\label{eq:F_fold}
    F(f) = \frac{2^{5/6}\pi^{1/2}}{i^{5/2}} \frac{w^{1/6}}{|\phi_{11}|^{1/2}|\phi_{222}|^{1/3}} \\
    \text{Ai}\left(\frac{2^{1/3} w^{2/3} y_2 }{|\phi_{222}|^{1/3}}\right)e^{-iw\frac{y_1^2}{2\phi_{11}}}\,,
\end{multline}
where $\mathrm{Ai}(x)$ is the Airy function of the first kind, a well-known mathematical function whose modulus has a fringe structure for $x<0$ and decays rapidly for $x>0$. 
This function has a maximum at $\approx 0.54$, and the maximum magnification $\mu_{\text{max}}(f)$ can be obtained from its modulus squared $|F(f)|^2$ at that point \cite{Schneider:1992bmb}. 
For a lens whose angular scale $\theta_{\star}$ is determined by a velocity dispersion $\sigma_v$, i.e., $\theta_{\star}=4\pi (D_{\text{LS}}/D_{\text{S}})(\sigma_v/c)^2$ and the lens-source distances are equipable, i.e., $D_{\text{L}} \sim D_{\text{LS}} \equiv D$, we find
\begin{multline}
    \left(\frac{\mu_\mathrm{max}}{10^3}\right) \approx \frac{3(1+\zl)^{1/3}}{2|1-\kappa|\,|\phi_{222}|^{2/3}}\left(\frac{f}{100\,\mathrm{Hz}}\right)^{1/3} \\ \left(\frac{D}{1\,\mathrm{Gpc}}\right)^{1/3}\left(\frac{\sigma_v}{200\,\mathrm{km\,s}^{-1}}\right)^{4/3}\,,
\end{multline}
where $\kappa$ is the local dimensionless surface mass density that is related to the second derivative of the Fermat potential, i.e., $|\phi_{11}| = 2|1-\kappa|$ \cite{Schneider:1992bmb}.
Therefore, for typical derivatives of the Fermat potential of order 1  and typical galaxy lenses, a lensed \gls{GW} signal at 250 Hz could only have a magnification up to a few thousand.

\begin{figure}
    \centering
    \includegraphics[width=\columnwidth]{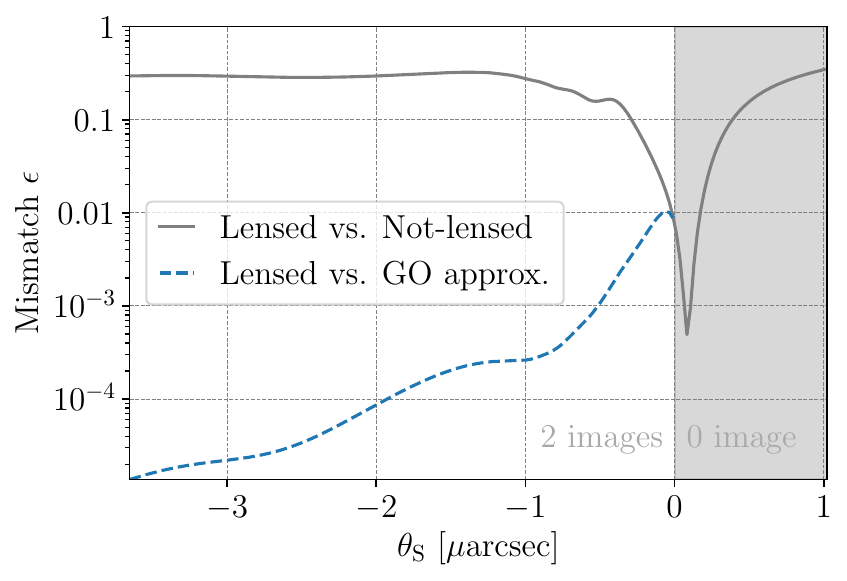}
    \caption{Mismatch between a lensed \gls{GW} and different waveform templates as a function of the angular separation to a fold caustic, $\theta_{\text{S}}$. 
    We compare the lensed signal for our fiducial $30M_{\odot} - 30M_{\odot}$ binary black hole merger waveform, computed using wave optics as in Eq.~\eqref{eq:F_fold}, with templates not including lensing (solid line) and templates using the \gls{GO} approximation in which two lensed images of different parity have a time delay, cf.~Eq.~\eqref{eq:F_two_images} (dashed line). 
    The \gls{GO} approximation matches well the lensed signal, which differs significantly from the not-lensed waveform due to the two images interfering. 
    In the shaded region there are (locally) no \gls{GO} images. 
    }
    \label{fig:comparison_wo_vs_go}
\end{figure}

To quantify the waveform distortion on a \gls{GW} signal due to diffraction near a fold caustic, we again compute the mismatch $\epsilon$ between a lensed signal and a not-lensed template.
The results are presented in Fig.~\ref{fig:comparison_wo_vs_go}, where we plot $\epsilon$ as a function of the angular source separation to the caustic $\theta_{\text{S}}$. 
Negative separations correspond to the higher multiplicity region (inside the caustic) with locally 2 images, while there are no local \gls{GO} images with positive separations (outside the caustic).
The mismatch is large, $\epsilon \approx 30\% $, for a large range of separations around the caustic (solid line).
This indicates that the lensed waveform is not properly described by an individual (not-lensed) template and that this effect is observable.
Interestingly, \gls{WO} predicts that there will be distortions even when the source is located outside a fold caustic.  
However, the amplitude of such highly distorted signals decays rapidly for large positive separations, $\theta_{\text{S}}\gg 1 \,\mu$arcsec in this example. %

We also quantify the difference between the analytic \gls{WO} solution around the caustic in Eq.~\eqref{eq:F_fold} and the \gls{GO} approximation in Eq.~\eqref{eq:F_two_images}, which is only valid for $\theta_{\text{S}}<0$. 
We find that the mismatch is generally small, indicating that the lensed signal is well described by the superposition of two \glspl{GW} with a small time delay and opposite parity.
Still, there could be a (comparatively) large mismatch between the two, which is maximal at the caustic (dashed line). This shows that diffraction signatures on the lensed \gls{GW} waveform are not fully captured by the \gls{GO} approximation.

Since the phase evolution of a \gls{GW} can be tracked over a large frequency range, in contrast to \gls{EM} signals whose intensity at a given frequency ($\propto|F|^2$) is typically the main observable and thus the information on the phase modulation from \gls{GL} is lost, \glspl{GW} are unique messengers to probe \gls{WO} distortions. We have shown that highly magnified \glspl{GW} around a fold caustic are perfect candidates to measure \gls{GL} diffraction for the \emph{first} time.

\lettersection{Conclusion}
Highly magnified \acrlongpl{GW} unlock the high-redshift Universe that would otherwise be unreachable by our detectors. %
Even with current-generation detectors, we are already sensitive to this kind of signals.
For example, with present sensitivities \cite{o4_sensitivity_curves_ligo}, \acrlongpl{GW} from a compact binary coalescence with total source-frame mass of $60M_{\odot}$ would be observable up to a redshift of $z \sim 20$, when the Universe was only 160 million years old and well beyond the usual (not-lensed) horizon of $z \sim 1$.
For lighter binary systems such as \acrlong{BNS} mergers, they could be observable up to a redshift of $z \sim 2$ already instead of the usual horizon of $z \sim 0.1$.
While the rate of observing these highly magnified signals is admittedly low, $\sim0.01\%$ of all events lensed by galaxies have $\mu>100$~\cite{Oguri:2018muv}, we find that they have very specific observational signatures that can be easily implemented and searched for in our data.

In particular, they present themselves as a pair of two signal-to-noise ratio peaks of roughly the same amplitude and closely separated in time, corresponding to the successive arrival of two lensed signals that are mirror reflected relative to the other.
These features are robust against one's choice in modeling the lenses.
Given the extremely low probability of such a chance coincidence due to phenomena other than lensing, no electromagnetic follow-up is needed in order to confirm this kind of lensing detection.
In extreme cases where geometric optics breaks down, diffraction becomes important. In fact, diffraction limits the maximum possible magnification and gives waveform distortions absent in geometric optics, where the gravitational-wave window is the only way to observe. %

\hfill\begin{acknowledgments}
\emph{Acknowledgments}.---We thank Amanda Farah and Aditya Vijaykumar for useful discussions about the detection rates of highly magnified gravitational waves and overlapping signals. 
This work was supported by Research Grants No.~VIL37766 and No.~VIL53101 from Villum Fonden and the DNRF Chair Program Grant No.~DNRF162 by the Danish National Research Foundation.
This work has received funding from the European Union's Horizon 2020 research and innovation programme under the Marie Sklodowska-Curie Grant Agreement No.~101131233.  
J.~M.~E is also supported by the Marie Sklodowska-Curie Grant Agreement No.~847523 INTERACTIONS.
The Center of Gravity is a Center of Excellence funded by the Danish National Research Foundation under Grant No.~184.
The Tycho supercomputer hosted at the SCIENCE HPC center at the University of Copenhagen was used for supporting this work.
\end{acknowledgments}

\bibliography{references}%

\end{document}

%% file: new_commands.tex
\newcommand{\ba}{\begin{eqnarray}}
\newcommand{\ea}{\end{eqnarray}}
\newcommand{\be}{\begin{equation}}
\newcommand{\ee}{\end{equation}}

\newcommand{\murel}{\mu_{\text{rel}}}

\newcommand{\zl}{z_{\text{L}}}
\newcommand{\zs}{z_{\text{S}}}

\definecolor{grey}{rgb}{0.4,0.4,0.4}
\definecolor{dullmagenta}{rgb}{0.4,0,0.4}
\definecolor{darkblue}{rgb}{0,0,0.4}
\definecolor{midblue}{rgb}{0,0,0.5}
\definecolor{midred}{rgb}{0.5,0,0}
\definecolor{orange}{rgb}{1,0.5,0}
\definecolor{lightbrown}{rgb}{0.75,0.5,0.25}
\definecolor{tan}{cmyk}{0.14,0.42,0.56,0}
\definecolor{djunglegreen}{cmyk}{0.99,0,0.52,0}
\definecolor{lightgreen}{rgb}{0,1,0}
\definecolor{olivegreen}{cmyk}{0.64,0,0.95,0.40}
\definecolor{midgreen}{rgb}{0.0,0.675,0.0}
\definecolor{darkgreen}{rgb}{0,0.5,0}